\documentclass[prl,twocolumn,aps,amssymb,footinbib]{revtex4}
\usepackage{amssymb}
\usepackage{graphicx}
\usepackage{amsmath}
\usepackage{times}
\usepackage{color}
\usepackage{subfigure}
\usepackage{setspace}
\usepackage{bm}

\newcommand {\beq} {\begin{equation}}
\newcommand {\eeq} {\end{equation}}
\newcommand {\bqa} {\begin{eqnarray}}
\newcommand {\eqa} {\end{eqnarray}}

\newcommand {\up} {\ensuremath{\uparrow}}
\newcommand {\dn} {\ensuremath{\downarrow}}

\newcommand{\cd} {\ensuremath{c^\dagger}}
\newcommand{\ca} {\ensuremath{c^{\phantom \dagger}}}
\newcommand{\kk} {\ensuremath{{\bf k}}}

\newcommand {\qq} {\ensuremath{{\bf q}}}
\newcommand {\rr} {\ensuremath{{\bf r}}}

\begin{document}
\begin{abstract}
  We study the quench dynamics of a two-component ultracold Fermi gas
  from the weak into the strong interaction regime, where the short
  time dynamics are governed by the exponential growth rate of
  unstable collective modes.  We obtain an effective interaction that
  takes into account both Pauli blocking and the energy dependence of
  the scattering amplitude near a Feshbach resonance. Using this interaction we
  analyze the competing instabilities towards Stoner ferromagnetism
  and pairing.
\end{abstract}

\title{Competition between pairing and ferromagnetic instabilities
in ultracold Fermi gases near Feshbach resonances}

\author{David Pekker$^1$, Mehrtash Babadi$^1$, Rajdeep Sensarma$^2$, Nikolaj Zinner$^{1,3}$, Lode Pollet$^1$, Martin W. Zwierlein$^{4}$, Eugene Demler$^1$}

\affiliation{
$^1$ Physics Department, Harvard University, Cambridge, Massachusetts 02138, USA\\
$^2$ Condensed Matter Theory Center, University of Maryland,
College Park, Maryland 20742, USA \\
$^3$ Department of Physics and Astronomy, Aarhus University, Aarhus, DK-8000, Denmark\\
$^4$ MIT-Harvard Center for Ultracold Atoms, Research Laboratory of Electronics,
and Department of Physics, Cambridge, MA 02139, USA}

\maketitle

Ferromagnetism in itinerant Fermions is a prime example of a strongly
interacting system. Most theoretical treatments rely on a mean-field
Stoner criterion~\cite{Stoner}, but whether this argument applies
beyond mean-field remains an open problem. It is known that the
existence of the Stoner instability is very sensitive to the details
of band structure and interactions~\cite{Kanamori, Tremblay, Tasaki},
however how to account for these details in realistic systems remains
poorly understood. Exploring the Stoner instability with ultracold
atoms has recently attracted considerable attention.  Following
theoretical proposals~\cite{UltracoldStoner}, the MIT group made use
of the tunability~\cite{Bloch:Review} and slow time
scales~\cite{SpinorBEC, Roberts2001, Greiner2002, Strohmaier:Decay} of
ultracold atom systems to study the Stoner
instability~\cite{MIT:Stoner:Expt}. Signatures compatible with
ferromagnetism, as understood from mean-field
theory~\cite{Paramekanti}, were observed in experiments: a maximum in
cloud size, a minimum in kinetic energy and a maximum in atomic losses
at the transition. However, no magnetic domains were resolved.

An important aspect of the MIT experiments is that they were done
dynamically: the Fermi gas was originally prepared with weak
interactions and then the interactions were ramped to the strongly
(repulsive) regime. Dynamic rather than adiabatic preparation was used
in order to avoid production of molecules.  This raises the
question of what are the dominant instabilities of the Fermi gas in
the vicinity of a Feshbach resonance.

Naively, one would expect that on the BEC-side, molecule production is
slow, as it requires a three-body process. Therefore, instability
towards Stoner ferromagnetism would dominate over the instability
toward molecule production. In this picture, one would expect that
quenches to the attractive (BCS) regime always yield an instability
towards pairing, whereas quenches to the repulsive (BEC) regime 
an instability towards ferromagnetism for sufficiently strong
interactions.

\begin{figure}
\includegraphics[width=8cm]{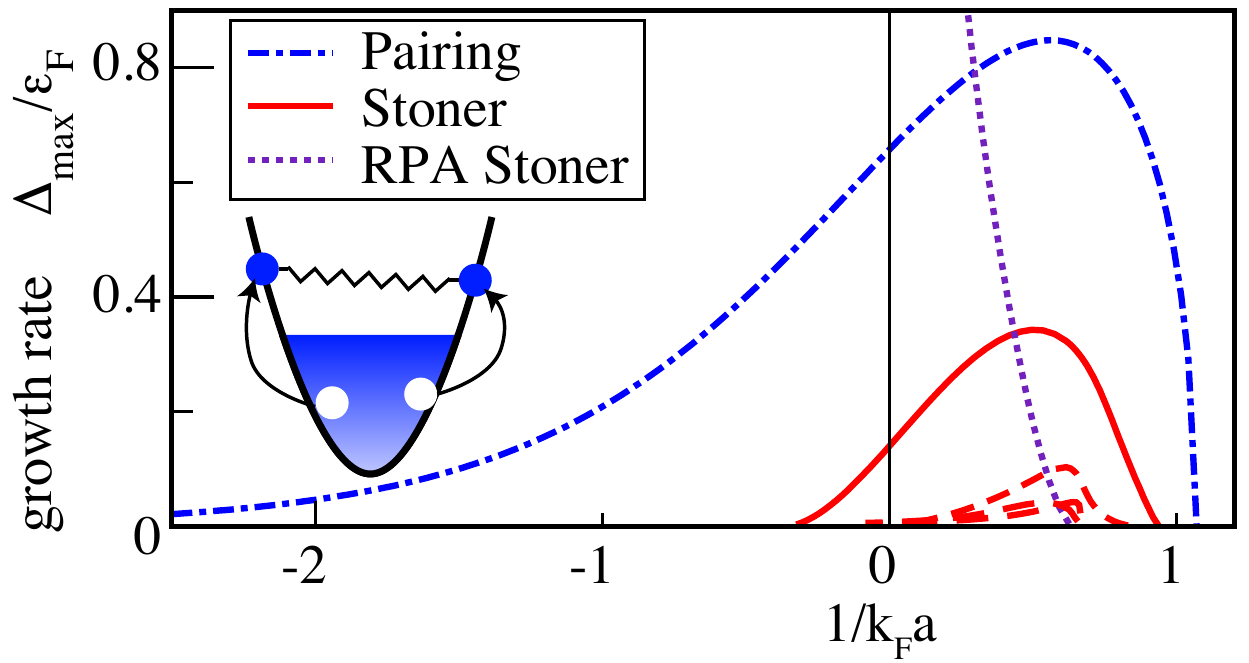}
\caption{Growth rate of the pairing and Stoner ferromagnetic
  instabilities after a quench as a function of the final interaction
  strength $1/k_Fa$. Final interactions with negative (positive)
  values of $1/k_Fa$ correspond to the BCS (BEC) side of the Feshbach
  resonance. The Stoner instability simultaneously occurs in multiple
  channels. The most unstable channel is indicated by the solid red
  line, the others by dashed red lines.  The ``RPA Stoner''
  instability corresponds to the RPA result with bare as opposed to
  Cooperon-mediated interaction (see text and Ref.~\cite{Babadi}).
  \newline {\bf Inset}: Schematic diagram of the pair creation process
  showing the binding energy (spring) being absorbed by the Fermi
  sea (arrows).}
\label{composite}
\end{figure}

In this Letter, we argue that this picture, which was used to
interpret the MIT experiments, is incomplete. Near the Feshbach
resonance, even on the BEC side, pair production remains a fast
two-body process as long as the Fermi sea can absorb the molecular
binding energy. As a result, near the Feshbach resonance, both on the
BEC and the BCS side, the pairing and the Stoner instabilities compete
directly. We now discuss these instabilities and their
competition in detail.

We start by describing the inter-atomic interactions. A Feshbach
resonance enables tunable interactions between ultracold atoms by
coupling the collision partners to a molecular state with different
magnetic moment. For broad resonances, where the coupling is much
larger than the Fermi energy, this can be modelled by a single
collision channel that supports one shallow bound
state~\cite{Varenna}.  An often used, but pathological choice, is to
describe repulsive interactions with a hard-sphere
pseudo-potential. Instead, one should use the full T-matrix that
includes the molecular bound state~\cite{Troyer}.  Although at low
energies the scattering amplitude from the hard-sphere potential and
the T-matrix match, at higher energies comparable to the molecular
binding energy, they do not. Specifically, in the strong interaction
regime where the Stoner instability is expected to occur, the Fermi
energy is comparable to the binding energy of a
molecule in vacuum, precluding the use of the hard-sphere potential.

In light of this remark, we study the initial dynamics of the
collective modes of a Fermionic system after a sudden quench, taking
the Cooperon (full T-matrix and Pauli blocking) into account.  We
focus on the case of a sudden quench, as it is simpler and captures
the essential physics of the instability of the Fermi
surface. Extensions to finite rate quenches are discussed in
Ref.~\cite{Babadi}.  Our main findings are summarized in
Fig.~\ref{composite} and are: (1)~We find that with the full T-matrix
the Stoner instability survives with a finite growth rate in the range
$-0.2 \lesssim k_F a \lesssim 1$, where $a$ is the scattering length
and $k_F$ is the Fermi momentum. In contrast, using bare
interactions~\cite{Babadi} results in an unphysical divergence of the
growth rate at unitarity and no magnetic instabilities on the BCS side
(see Fig.~\ref{composite}). (2)~The pairing instability persists on
the BEC side, where it competes with the Stoner
instability. (3)~Within our approximations, the pairing instability is
stronger.

At first sight, the survival of the pairing and Stoner instabilities
on the wrong side of the resonance is quite remarkable.  However, both
can be understood by taking into account the presence of the Fermi
sea. On the BEC side, due to Pauli blocking, the binding energy of the
pair-like molecule can be absorbed by the two holes that are left
behind (see the inset of Fig.~\ref{composite}).  Thus, the two-body
pairing process becomes forbidden when the binding energy $\sim 1/m
a^2$ exceeds the maximum energy that can be absorbed by the holes
$\sim k_F^2/m$ ($m$ is the Fermion mass, $a$ the scattering length,
$k_F$ the Fermi momentum, and throughout this Letter we use the units
in which $\hbar=1$).  On the BCS side, although interactions at low
energies are indeed attractive, the same is not true at high
energies. As the Stoner instability involves all scattering energies
up to the Fermi energy, it is natural that it can persist around
unitarity, even on the BCS side.

{\it Formalism --\/ } We consider a system of interacting Fermions
described by the Hamiltonian:
\begin{align}
\!\!H=\sum_{\kk,\sigma} \xi_{\kk \sigma} \cd_{\kk\sigma} \ca_{\kk\sigma}  \!+\!
\int d^d\rr \, U(t,\rr-\rr') \cd_{\rr \up} \cd_{\rr' \dn} \ca_{\rr' \dn} \ca_{\rr \up},
\label{eq:H}
\end{align}
where $\cd_\sigma (\ca_\sigma)$ are the Fermion creation
(annihilation) operators with spin $\sigma$, $\xi_{\kk
  \sigma}=k^2/2m-\mu_\sigma$, $\mu_\sigma$ are the chemical
potentials, and $U(t,\rr-\rr')$ is the time dependent pseudo-potential
that describes the inter-atomic interaction. We focus on the
instantaneous quench limit, in which the coupling $U$ changes from a
negligible initial value $U_i$ to a final value $U_f$ at time $t=0$.
In this limit, we can describe short time dynamics of a collective
mode at momentum $q$ using the corresponding susceptibility,
$\chi_\qq(\omega_\qq;U_f)$, evaluated with final interactions but
initial Fermionic configuration~\cite{Lamacraft, Babadi}. In
particular, if $\chi_\qq(\omega_\qq;U_f)$ has poles at
$\omega_\qq=\Omega_\qq+ i \Delta_\qq$ in the upper half of the complex
plane, then fluctuations that occur after the quench will grow
exponentially in time. Next, we obtain a universal description of
interactions created by the pseudo-potential $U(\rr)$ by modifying the
T-matrix formalism to take into account Pauli blocking, and apply
these ideas to the Stoner and BCS instabilities.

\begin{figure}
\includegraphics[width=9cm]{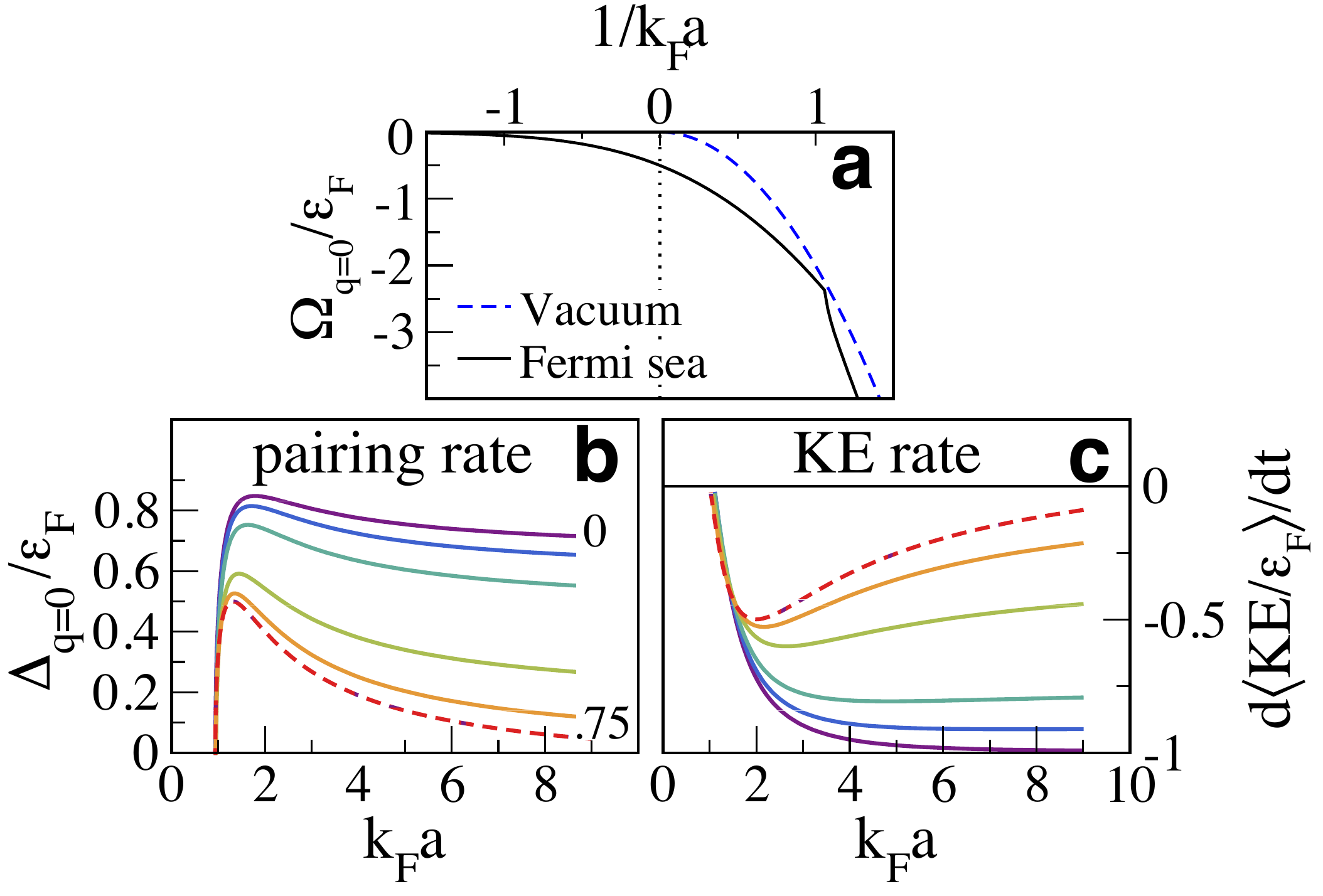}
\caption{Pairing instability. (a) ``Binding energy'' of a Feshbach
  molecule in vacuum and in the presence of a Fermi sea (relative to
  $2\epsilon_F$) as a function of interaction strength, corresponding
  to the real part of the T-matrix pole frequency
  $\Omega_{q=0}=\text{Re}[\omega_{q=0}]$. Pauli blocking by the Fermi
  sea results in stronger binding across the resonance. The kink
  occurs when the pair becomes stable. (b)~Pairing rate and (c)~rate
  of change of Kinetic Energy as a function of interaction strength on
  the BEC side for various temperatures [$T=0$~(purple, solid), 0.12,
  0.22, 0.5, 0.66, $0.75 \, T_F$~(red, dashed)].  Temperature is more
  effective at suppressing pair production at larger values of $k_f a$
  as the binding energy is smaller, thus the peaks in (b) and (c)
  become sharper at higher temperatures. 
  The peaks in growth rate and kinetic energy rate qualitatively match
  experiments~\cite{MIT:Stoner:Expt}. Sharp onset at $k_F a
  \approx 1.1$ is expected to be smoothed by three-body processes.  }
\label{imbalance}
\end{figure}

\begin{figure}
\includegraphics[width=6.5cm]{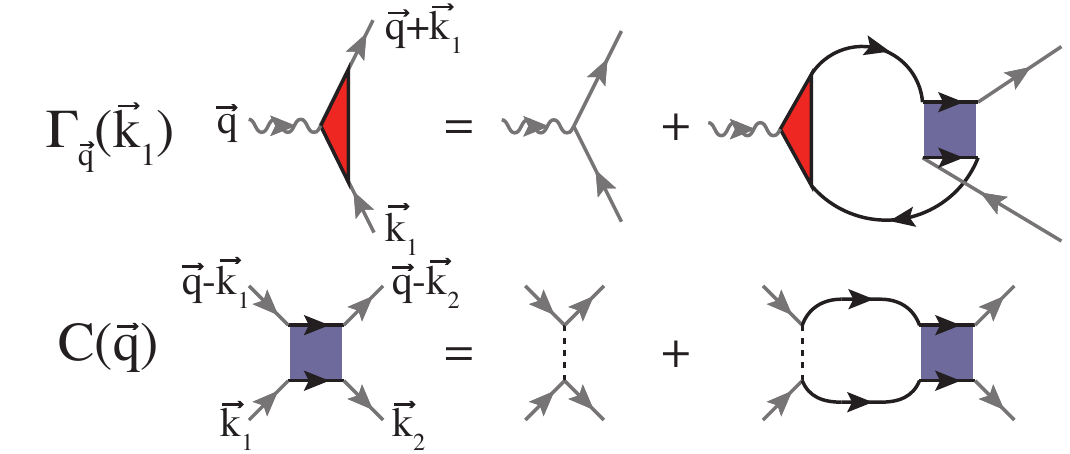}
\caption{Diagrams for the vertex function $\Gamma$ and Cooperon $C$.
  Solid lines represent bare fermion propagators, dashed lines 
  interactions, and wavy lines external sources. External legs, represented
  by gray lines, are shown for clarity.  }
\label{fig:diagrams}
\end{figure}

{\it Cooperon --\/} In this section, we obtain the
Cooperon, $C$, i.e.  the T-matrix that takes into account Pauli
blocking of states by the Fermi sea (see Fig.~\ref{fig:diagrams}),
which is needed for an accurate study of instabilities near a Feshbach
resonance.

In the center of mass frame in vacuum, the scattering of a pair of
particles with identical masses $m$ near a wide Feshbach resonance is
described by the T-matrix (scattering amplitude)
\begin{equation}
\tau(E)=\frac{m}{4 \pi}\left(\frac{1}{a} +i\sqrt{m E}\right)^{-1}.
\end{equation}
Here, $E$ is the energy of the scattered particles and the
pseudo-potential $U(\rr-\rr')$ that appears in Eq.~\ref{eq:H} is
related to the T-matrix via the Lippmann-Schwinger equation. To
correctly renormalize the Cooperon, we compare the Lippmann-Schwinger
equations in a Fermi sea and in vacuum. For the Cooperon we can not
just use the center of mass frame, as the Fermi sea breaks
translational invariance. Therefore, we use the laboratory frame for
both to obtain
\begin{align}
C^{-1}(E,\qq)&=\tau^{-1}\left(E+2\epsilon_f-\qq^2/4 m\right)\nonumber\\
&\quad\quad+
\int \frac{d^3\kk}{(2\pi)^3} \frac{n^F(\frac{\qq}{2}+\kk)+n^F(\frac{\qq}{2}-\kk)}{E -\epsilon_{\frac{\qq}{2}+\kk}-\epsilon_{\frac{\qq}{2}-\kk}}.
\label{eq:cooperon}
\end{align}
Here, $E$ and $\qq$ are the center of mass frequency and momentum of
the pair, $\epsilon_f$ is the Fermi energy, $n^F(\kk)$ is the Fermi
function, and $\epsilon_{\kk}=\kk^2/2m-\epsilon_f$. Our approach is
analogous to the one used for the Fermi-polaron
problem~\cite{Prokofev2008}.

{\it Pairing instability --\/} The Cooperon enters the
pairing susceptibility via
\begin{align}
\chi_\text{pair}(\vec{q})&=\int d\vec{k}_1 \, d\vec{k}_2 \, G(\vec{k}_1) G(\vec{q}\!-\!\vec{k}_1) C(\vec{q})\times \nonumber\\
&\quad\quad\quad\quad\quad\quad\quad\quad\quad\times G(\vec{k}_2) G(\vec{q}\!-\!\vec{k}_2),
\end{align}
where $\vec{q}$ stands for the external frequency and momentum vector
$\{E,\qq\}$, $d\vec{k}_1$ stands for $d\omega_1\, d\kk_1/(2\pi)^4$,
and $G(\vec{k}_1)=G(\omega_1,\kk_1)$ is the bare Fermionic Green
function in the non-interacting Fermi sea corresponding to the initial
state. The poles of the pairing susceptibility correspond thus to the
poles of the Cooperon, whose structure we now explain.

We begin our analysis with the T-matrix in vacuum. For each value of
the scattering length, the T-matrix has a line of poles on the BEC
side located at $\omega_q=\Omega_q+i\Delta_q=-1/ma^2+m q^2/4$,
corresponding to the binding energy of a Feshbach molecule with center
of mass momentum $q$. As a consequence of energy and momentum
conservation the pole frequency is strictly real ($\Delta_q=0$),
indicating that a two-body process in vacuum cannot produce a Feshbach
molecule.

In the presence of a Fermi sea, the states below the Fermi surface are
Pauli-blocked, shifting the poles of the Cooperon relative to the
T-matrix in vacuum in two important ways. First, the real part of the
pole $\Omega_q$, which would correspond to the binding energy of a
pair in the absence of an imaginary part, uniformly shifts down (see
Fig.~\ref{imbalance}a). This shift is a result of Pauli
blocking~\cite{AGD}, and indicates an appearance of a paired state
on the BCS side as well as stronger binding of the pair on the BEC
side.  Second, in the range $-\infty<1/k_F a\lesssim 1.1$ the pole
acquires a positive imaginary part $\Delta_q$ that corresponds to the
growth rate of the pairing instability. As depicted in
Fig.~\ref{composite}, the growth rate of the pairing instability
increases exponentially $\Delta_{q=0} \approx 8 \epsilon_F e^{\pi/2
  k_F a-2}$ as one approaches the Feshbach resonance from the BCS
side, {\it i.e.} the growth rate of the BCS pairing is equal to the
BCS gap at equilibrium~\cite{AGD}. On the BEC side, the growth rate
continues to increase, reaching a maximum at $k_F a \approx 2$, and
finally decreasing to zero at $k_F a \approx 1.1$, at which point the Fermi
sea can no longer absorb the energy of the Feshbach molecule in a
two-body process. Pairing deeper in the BEC regime takes place via the
more conventional three-body process and would round the pairing
instability curve near $k_F a \approx 1.1$ in Fig.~\ref{composite}.

So far, we have concentrated on the instability at $q=0$. We find that
$q=0$ is indeed the most unstable wavevector throughout the Feshbach
resonance, but the growth rate remains finite up to
$q=q_\text{cut}$. Throughout the resonance the approximation
$q_\text{cut} \approx (\sqrt{3/2}) (\Delta_{q=0}/\epsilon_F) k_F$
works reasonably well except in the vicinity of $k_f a \sim 2$ where
$q_\text{cut}$ reaches the maximal value for a two-body process of $2 k_f$.

\begin{figure}
\includegraphics[width=8.5cm]{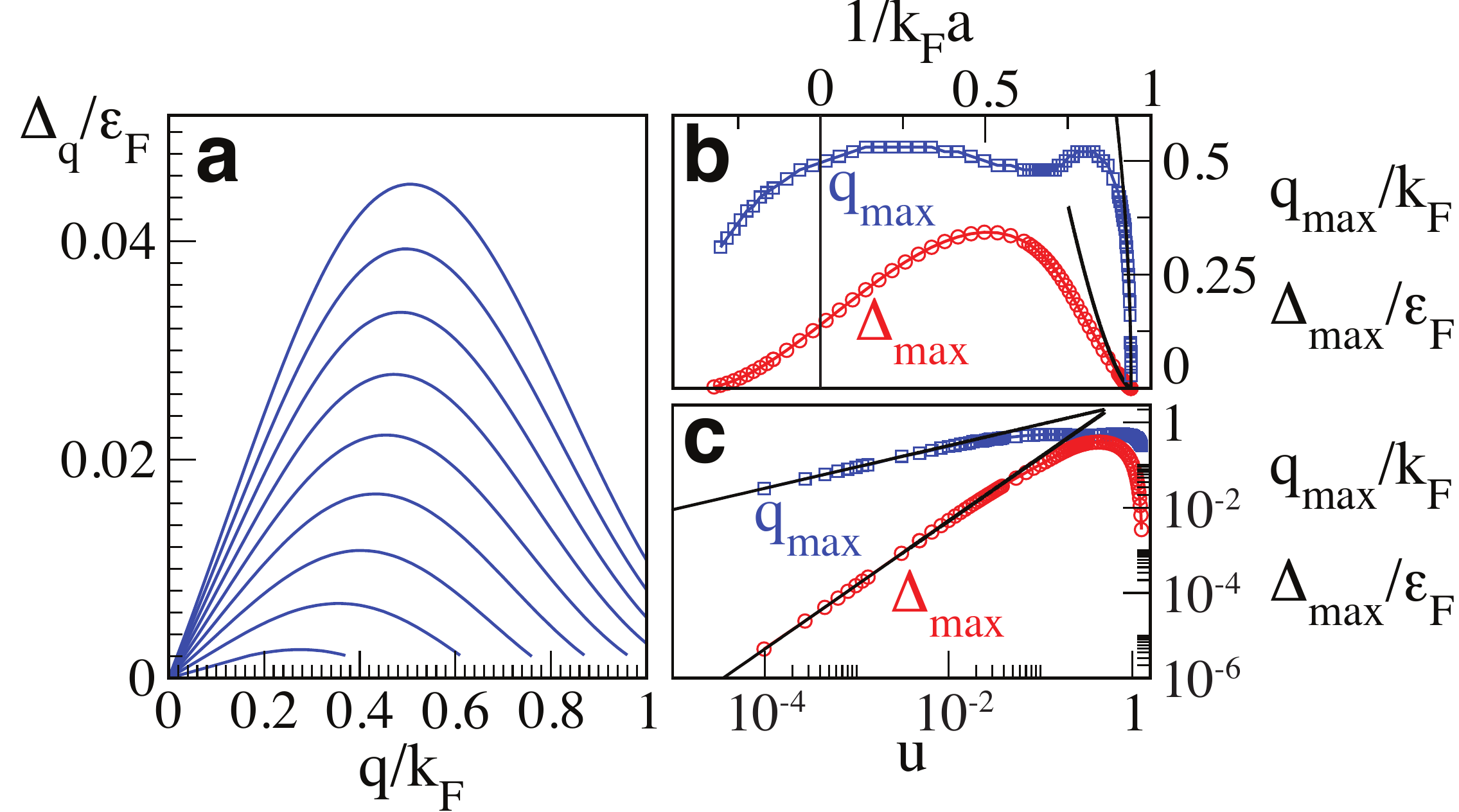}
\caption{Properties of growing collective modes in the Stoner
  instability in 3D. (a)~Growth rate of the most unstable mode
  $\Delta_q$ as a function of wavevector $q$ for $T=0$ and $1/k_F
  a=0.85$~(top line), $0.86$, $0.87$, ..., $0.93$~(bottom line).
  (b)~The most unstable wavevector $q_\text{max}$ (blue) and the
  corresponding growth rate $\Delta_\text{max}$ (red) vs. $1/k_F a$.
  A fit to the mean-field critical theory ($\nu=1/2$, $z=3$) is shown
  with black lines~\cite{Vojta}. (c)~Details of the critical behavior of
  $q_\text{max}$ and $\Delta_\text{max}$ as a function of distance
  from the transition point $u=(1/k_Fa)_c-(1/k_Fa)$,
  $(1/k_Fa)_c\approx0.94$. }
\label{fig:stoner}
\end{figure}

{\it Stoner instability --\/} One can expect that a rapid quench to
the BEC side of the resonance, where interactions are strongly
repulsive, results in an instability towards Stoner ferromagnetism.  We
shall assume that right after the quench, the atoms are still in the
free Fermi sea initial state and the Stoner instability is competing
with the pairing instability. Our goal is to compute the ferromagnetic
susceptibility using the Cooperon to describe effective inter-atomic
interactions. Using the full Cooperon allows us to include three
important aspects of the problem: energy dependence of the scattering
amplitude near the Feshbach resonance; Pauli blocking, which
renormalizes the energy of the virtual two particle bound states
involved in scattering; and Kanamori-like many-body
screening~\cite{Tremblay}.

Technically, we compute the vertex function
$\Gamma_{\omega,\qq}(\omega_1,\kk_1)$, which is related to the
susceptibility via
\begin{align}
\chi_\text{FM} (\vec{q})=\int d\vec{k}_1 \, G(\vec{q}+\vec{k}_1) G(\vec{k}_1) \, \Gamma_{\vec{q}} (\vec{k}_1).
\end{align}
We note, that the poles of the susceptibility and the vertex function
coincide. Replacing the point contact interaction vertex by the
Cooperon in an RPA type resummation of the vertex function (see
Fig.~\ref{fig:diagrams} and Ref.~\cite{Tremblay}) we obtain 
\begin{align}
  &\Gamma_{\vec{q}}(\vec{k}_1)=1\nonumber\\&\quad+\int d\vec{k}_2 \,
  \Gamma_{\vec{q}}(\vec{k}_2) C(\vec{k}_1+\vec{k}_2+\vec{q})
  G(\vec{k}_2+\vec{q}) G(\vec{k}_2).
\label{Gamma-eq}
\end{align}

To compute the vertex function, a number of approximations are unavoidable.
First, we assume that $\qq$ and $\omega$ are both small, which is
valid in the vicinity of the Stoner transition. Second, in the spirit
of Fermi liquid theory, we assume that the most important poles come
from the Green functions, and thus we replace $G(\kk_2+\qq,
\omega_2+\omega)G(\kk_2,\omega)\rightarrow \frac{2 \pi}{v_F} \frac{\qq
  \cdot \kk_2}{m\, \omega-\qq \cdot \kk_2} \delta(\omega)
\delta(|\kk_2|-k_F)$~\cite{AGD}. We then obtain
\begin{align}
\!\!\Gamma_{\qq, \omega}(\hat{\kk}_1)\!=\!1\!+\!\int \frac{d\hat{\kk}_2}{4\pi} \Gamma_{\qq, \omega}(\hat{\kk}_2)C(\hat{\kk}_1\!+\!\hat{\kk}_2,\omega) I_{\qq,\omega}(\hat{\kk}_2),
\end{align}
where
\begin{align}
I_{\qq,\omega}(\hat{\kk}_2)=\int \frac{k_2^2 dk_2}{2\pi^2}\frac{n_F(\kk_2-\qq/2)-n_F(\kk_2+\qq/2)}{\omega - \epsilon_{\kk_2-\qq/2}+\epsilon_{\kk_2+\qq/2}},
\end{align}
and $\hat{\kk}$ indicates a vector on the Fermi surface.  We thus make
the approximation that we can replace $\kk_1$ and $\kk_2$ by
$\hat{\kk}_1$ and $\hat{\kk}_2$ when we evaluate the value of the
Cooperon. In other words, we assume that the Cooperon changes slowly
compared to the Green functions. The approximation is fully justified
for weak interactions, where the Cooperon is momentum and frequency
independent, and the vertex function matches the RPA
result~\cite{Babadi}.  For strong interactions, the Stoner instability
is not driven by the pole of the Cooperon, and we therefore believe
that our approximation captures the essential physics.

In the range $ - 0.2 \lesssim 1/k_F a \lesssim 1.0$, there is one or
more lines of complex poles with a positive imaginary part
$\Delta_\qq$, which corresponds to the Stoner instability in different
channels (a combination of momentum and orbital moment).  As
$q\rightarrow 0$, the different instabilities can be identified as
different angular momentum channels.  Since magnetization is a
conserved order parameter, in each channel $\Delta_{\qq}$ grows
linearly for small $q$. At large $q$ the cost of bending the order
parameter results in the vanishing of $\Delta_{\qq}$ for
$q>q_{\mathrm{cut}}$. In between, $\Delta_{\qq}$ reaches its maximum
value $\Delta_{\mathrm{max}}$ at a wave-vector $q_{\mathrm{max}}$
which corresponds to the fastest growing mode (see
Fig.~\ref{fig:stoner}).

{\it Discussion --\/} The growth rates of the pairing instability
$\Delta^\text{BCS}_{q=0}$ and the ferromagnetic instabilities in the
various channels $\Delta^\text{FM}_{\mathrm{max}}$ are compared across
the Feshbach resonance in Fig.~\ref{composite}. We have also included
the naive RPA estimate of the growth rate of the Stoner instability in
which we have replaced the Cooperon by $4\pi a/m$. From the
comparison, we see that (1) the Cooperon suppresses the growth rate of
the ferromagnetic instability but does not eliminate it, (2) the
pairing and ferromagnetic instabilities compete over a wide range of
interaction strength on both sides of the resonance, and (3) the
pairing instability is always dominant. Our results suggest that even
if there is a metastable ferromagnetic state~\cite{Troyer}, it
probably cannot be reached via dynamic tuning of the interaction
starting from a balanced gas. However, for short timescales $\sim
\left(\Delta^\text{FM}_{\mathrm{max}}\right)^{-1}\sim\left(\Delta^\text{BCS}_{q=0}\right)^{-1}$,
both pairing and magnetic correlations will develop
and may be detectable experimentally.

{\it Comparison with experiment --\/} The maximum of the pairing
instability in the vicinity of $k_F a \approx 2$ closely matches the
location of the transition found
experimentally~\cite{MIT:Stoner:Expt}. To investigate further, we plot
the pairing rate as a function of interaction strength for several
different temperatures in Fig.~\ref{imbalance}b. The shape of the
pairing rate curve, especially at higher temperatures looks
qualitatively similar to the atom loss rate (into pairs)
found experimentally. 

A fast rampdown of the magnetic field was used to convert weakly bound
molecules into strongly bound molecules, and the kinetic energy of the
remaining atoms was measured. It was found to have a minimum at $k_F
a\approx 2$~\cite{MIT:Stoner:Expt}. We show that this minimum can be
qualitatively understood within our analysis of the pairing
instability. The energy of each molecule produced is given by $\sim
\text{Re}[\omega_q]$ (see Fig.~\ref{imbalance}a). The molecular energy
corresponds to the kinetic energy of the Fermions removed from the
Fermi sea, measured with respect to the Fermi energy. Thus the rate of
kinetic energy change of ``unpaired'' atoms is $\sim
(\text{Re}[\omega_q]-2\epsilon_F) \times \text{Im}[\omega_{q=0}]$ (see
Fig.~\ref{imbalance}c). We find that the kinetic energy minimum is in
the vicinity of the maximum of the pairing rate, in qualitative
agreement with Ref.~\cite{MIT:Stoner:Expt}.

{\it Acknowledgements --\/ } It is our pleasure to thank E. Altman,
A. Chubukov, D. Huse, M. Lukin, S. Stringari, I. Carusotto, A. Georges
and especially W. Ketterle for useful discussions. The authors
acknowledge support from a grant from the Army Research Office with
funding from the DARPA OLE program, CUA, the Swiss national Science
Foundation, NSF Grant No. DMR-07-05472 NSF and PHY-06-53514,
AFOSR-MURI, and the Alfred P. Sloan Foundation.

\end{document}